\documentclass[usenatbib]{mn2e}

\usepackage{natbib} 
\usepackage{graphicx}
\usepackage{psfrag}

\def\vec#1{\ensuremath{\mathchoice{\mbox{\boldmath$\displaystyle#1$}}
{\mbox{\boldmath$\textstyle#1$}}
{\mbox{\boldmath$\scriptstyle#1$}}
{\mbox{\boldmath$\scriptscriptstyle#1$}}}}

\def\O{{\cal O}}
\def\text#1{{\rm #1}}

\def\mb{\mbox}
\def\qaq{{\quad\text{and}\quad}}

\def\X{{\mathit{X}}}
\def\Y{{\mathit{Y}}}

\def\Lagr{{\Lambda}}

\def\d{\delta}

\def\qaq{{\quad\text{and}\quad}}
\def\E{{\mathcal{E}}}

\def\Dv{\Delta}
\def\entr{\alpha}

\def\vxi{ {\vec{\xi}} }

\def\n{{\rm n}}
\def\p{{\rm p}}

\def\b{\mathrm{b}}

\def\implies{{\quad\Longrightarrow\quad}}

\def\eps{\varepsilon}
\def\epsn{\eps_\n}
\def\epsp{\eps_\p}

\def\vn{{\vec{n}}}
\def\vv{{\vec{v}}}
\def\vnabla{{{\nabla}}}
\def\vDv{{\vec{\Dv}}}

\def\vp{\vec{p}}

\def\csum{\sum}

\def\vf{\vec{f}}

\def\mutual{{\mathrm{mut}}}

\def\mut{\widetilde{\mu}}

\def\Lie{\pounds}

\def\vS{{\vec{S}}}

\def\1{{(1)}}
\def\2{{(2)}}

\def\Om{\Omega}
\def\om{\omega}
\def\omh{\widehat{\om}}
\def\mn{m_\b}
\def\ph{\varphi}
\def\vph{\vec{\ph}}
\def\th{\theta}
\def\til{\widetilde}
\def\vz{\hat{\vec{z}}}

\def\Cor{\mathcal{C}}
\def\vCor{\vec{\Cor}}

\def\vR{\vec{R}}
\def\vS{\vec{S}}
\def\vT{\vec{T}}

\def\kap{\kappa}

\def\xp{x_\p}
\def\xn{x_\n}

\def\R{\mathcal{R}}
\def\psih{\widehat{\psi}}

\def\gam{\gamma}
\def\gamn{\nu_\n}
\def\gamp{\nu_\p}
\def\gamX{\nu_\X}

\def\cs{{c_\textrm{s}}}
\def\co{{c_0}}

\def\sf{{\mathrm{sf}}}
\def\ord{{\mathrm{ord}}}
\psfrag{1I1}{\hspace*{-0.5cm}$^1_2I_{(1)}$ (r-mode)}
\psfrag{2I1}{$^2_2I_{(1)}$}
\psfrag{2I2}{$^2_2I_{(2)}$}
\psfrag{2I12}{\hspace*{-0.5cm}$^2_2I_{(1)}$ and $^2_2I_{(2)}$}
\psfrag{3I1}{$^3_2I_{(1)}$}
\psfrag{3I2}{$^3_2I_{(2)}$}
\psfrag{3I3}{$^3_2I_{(3)}$}
\psfrag{3I123}{\hspace*{-1cm}$^3_2I_{(1)}$, $^3_2I_{(2)}$ and $^3_2I_{(3)}$}
\psfrag{Rel}{$\R$}
\psfrag{k0}{$\kap_0$}

\psfrag{kord}{$\kap_\ord$}
\psfrag{ksf}{$\kap_\sf$}
\psfrag{k+}{$\kap_+$}
\psfrag{k-}{$\kap_-$}
\psfrag{Imk}{$\hspace*{-0.1cm}\mathrm{Im}(\kap)$}
\psfrag{Rek}{$\hspace*{-0.1cm}\mathrm{Re}(\kap)$}
\psfrag{eps}{$\eps$}

\begin{document}

\title{Inertial modes of non-stratified superfluid neutron stars.}

\author[R.~Prix, G.L.~Comer and N.~Andersson]
{R.~Prix$^\dag$, G.L.~Comer$^*$ and N.~Andersson$^\dag$\\
$^\dag$Department of Mathematics, University of Southampton, Southampton, SO17 1BJ, United 
Kingdom \\$^*$Department of Physics, Saint 
Louis University, St Louis, MO 63156,  USA}

\maketitle

\begin{abstract}
We present results concerning adiabatic inertial-mode oscillations
of non-stratified superfluid neutron stars in Newtonian gravity, using
the anelastic and slow-rotation approximations. We consider a simple
two-fluid model of a superfluid 
neutron star, where one fluid consists of the superfluid neutrons
and the second fluid contains all the comoving constituents
(protons, electrons). The two fluids are assumed to be ``free'' in the
sense that vortex-mediated forces like mutual friction or
pinning are absent, but they can be \emph{coupled} by the equation of state, in
particular by entrainment.  The stationary background consists of the
two fluids rotating uniformly around the same axis with potentially different 
rotation rates.
We study  the special cases of co-rotating backgrounds, 
vanishing entrainment, and the purely toroidal r-modes, analytically. 
We calculate numerically the eigenfunctions and frequencies of inertial
modes in the general case of non co-rotating backgrounds, and study
their dependence on the relative rotation rate and entrainment. In
these non-stratified models we find  avoided crossings only between 
associated mode-pairs, eg. an
``ordinary'' mode and its ``superfluid'' counterpart, while other
mode-frequencies generally cross as the background parameters are varied.
We confirm (for the first time in a mode calculation) the onset of a ``two-stream
instability''  at a critical relative background 
rotation rate, and study some of the 
properties of this instability for the inertial modes.
\end{abstract}


\section{Introduction}

The oscillations of rotating compact stars is a subject that 
has attracted interest for a considerable time. This is natural
since the associated issues range from fundamental applied mathematics
(eg. the stability of rotating self-gravitating fluid configurations),
to mainstream astrophysics (eg. Helioseismology and attempts to 
infer the Sun's rotation profile from observed modes of oscillation)
and exotic neutron star physics (eg. the gravitational-wave driven instability
of the r-modes and various viscous damping mechanisms, like hyperon 
bulk viscosity). To date, most investigations have assumed that a rotating star can be
appropriately described by a perfect fluid model. While such models 
are relevant in many contexts, they do not provide an adequate 
description of mature neutron stars. Once a neutron star has cooled below 
$10^9-10^{10}$~K, i.e. within minutes to months after its birth, it's outer layers
will form a crystalline lattice of nuclei. At the same time, the fluid core is 
expected to contain several superfluid/superconducting components. 

This paper concerns the dynamics of rotating superfluid neutron stars. 
In particular, we study the inertial modes of a simple two-fluid model 
appropriate for the conditions that prevail in the outer core of a neutron star. 
The two fluids, which represent superfluid neutrons and a congomerate
of all  comoving constituents
(protons, electrons), are coupled via the equation of state
(in particular via entrainment), but are otherwise allowed to move at
independent velocities. Our background model is a stationary two-fluid
configuration with constant entrainment, with the two fluids rotating
uniformly around the same axis with rotation rates $\Om_\n$ and $\Om_\p$.  

Previous studies of superfluid inertial modes (including preliminary 
studies of the zero-frequency subspace), namely
\citet{lindblom00:_r_modes_superfl_neutr_stars,sedrakian00:_perturb_sf_normal_mixtures,andersson01:_dyn_superfl_ns,comer02:_zero_freq_subspace,lee03:_superfl_r_modes,yoshida03:_sf_inertial_modes,yoshida03:_r-modes_relat_superfl},
have all been restricted to co-rotating backgrounds
\mb{$\Om_\n=\Om_\p$}. This study is the first to allow for
the general case of a  background with two fluids rotating at
different rates. This is expected to be the quasi-stationary
``ground state'' of a superfluid neutron star due to its
emission-induced spindown and the weak coupling to the superfluid
components. This background model is also the starting point of all
viable models of Vela-sized glitches. 
In other words, by allowing for different rates of rotation, we have taken a crucial step towards more
realistic modelling of the dynamics of mature neutron stars.

\section{The two-fluid neutron star model}

We take as our starting point the ``standard'' 
two-fluid model for superfluid neutron
stars
(e.g. \cite{lindblom94:_oscil_superfl_ns,lee95:_nonrad_osc_superfl_ns,langlois98:_differ_rotat_superfl_ns,prix02:_adiab}),
in which one assumes that the protons and electrons are locked together
by the magnetic field and viscosity, while the superfluid neutrons
form an independent fluid due to their lack of viscosity. 
Our model neglects the presence of the elastic crust as well as the potential presence of
exotic matter in the deep neutron star core. In essence, the model
is expected to be relevant for the outer neutron star core. By studying the global modes
of oscillation of this model, we hope to gain insight into the complex dynamics 
of any two-fluid system. Even though we are not considering
a detailed realistic neutron star model (the construction of which would be very 
difficult given our current level of understanding)
we expect to learn much about qualitative aspects that should remain relevant also 
in more complicated settings. It is also interesting to note,  
cf. comments made by
\citet{sedrakian00:_perturb_sf_normal_mixtures}, that the study of
two-fluid models  may be of significance in laboratory contexts, for
example in the study of rotating heavy nuclei using the compressible
liquid approximation of the Bohr-Wheeler model, or for rotating
mixtures of Bose-Einstein condensates.  

A general Newtonian formalism to describe mixtures of charged and
uncharged fluids has been developed by 
\citet{prix02:_variat_I,prix02:_variat_II}, based on a
variational principle that was first developed in a fully
relativistic framework by Carter and coworkers 
\citep{carter89:_covar_theor_conduc,carter92:_momen_vortic_helic,carter98:_relat_supercond_superfl}.
In particular,  \cite{prix02:_variat_I} developed a general two-fluid neutron star model 
allowing for temperature gradients and dissipation through
mutual friction and $\beta$-reactions between the two fluids.
For the present application, however, we assume a ``cold'' neutron
star in which we can neglect temperature effects, so we set $T=0$, and
we also neglect mutual friction and non-adiabatic processes like
$\beta$-reactions. The resulting framework, which is identical to that 
used by \cite{prix02:_adiab}
and \cite{andersson01:_dyn_superfl_ns},
is briefly introduced in this section.  
Note that although the formalism used here is different from
the one more commonly found in the Newtonian literature (e.g.
\cite{lindblom94:_oscil_superfl_ns,lee95:_nonrad_osc_superfl_ns,lee03:_superfl_r_modes}),
which is based on the ``orthodox''
superfluid formalism introduced by Landau, the two frameworks can be
shown to be strictly equivalent, as discussed by \cite{prix02:_variat_I}.

Our two-fluid model consists of a neutron and a ``proton'' fluid
(the latter actually consists of the comoving protons \emph{and} the electrons). 
Therefore the kinematic variables are the particle number densities
$n_\n$ and $n_\p$ together with the respective transport velocities
$\vv_\n$ and $\vv_\p$. The corresponding transport currents are
naturally expressed as
\begin{equation}
\vn_\X = n_\X\, \vv_\X\,,
\end{equation}
where \mb{$\X \in \{\n,\,\p\}$} is the constituent index (the repetition of which does not 
imply summation).
An important quantity for our analysis
is the relative velocity $\vDv$ between the two
fluids, which we define as
\begin{equation}
\vDv \equiv \vv_\p - \vv_\n\,.
\end{equation}
The dynamics is governed by the Lagrangian density 
\begin{equation}
\Lagr = {1\over2} n_\n m_\n \vv_\n^2 + {1\over 2} n_\p m_\p \vv_\p^2
- \E - \rho\,\Phi\,,
\label{eq:DefLagr}
\end{equation}
where $\rho\equiv m_\n n_\n + m_\p n_\p$ is the total mass density,
$\Phi$ is the gravitational potential and $\E$ the energy function or
``equation of state'' of the system. The general form of the equation
of state is \mb{$\E=\E(n_\n, n_\p, \vDv^2)$}, which determines the
first law of thermodynamics in the form  
\begin{equation}
d\E = \mu^\n \, d n_\n + \mu^\p \, d n_\p + \entr\, d \vDv^2\,,
\label{eq:FirstLaw}
\end{equation}
defining the chemical potentials $\mu^\n$ and $\mu^\p$, as well as the
entrainment $\entr$. 
The conjugate momenta for the two fluids are defined by the total
differential of the Lagrangian density $\Lagr$, namely
\begin{equation}
d\Lagr = \csum_{\X=\n,\p}\left[ \vp^\X \cdot d\vn_\X + (p_0^\X - m^\X\Phi)\,d
  n_\X\right] - \rho \, d\Phi\,, 
\end{equation}
In the following we assume the two masses to be equal, so we set
$m^\p = m^\n= m_\b$.  
With the explicit form (\ref{eq:DefLagr}) of the Lagrangian and the
first law (\ref{eq:FirstLaw}), we can express these conjugate momenta
 as 
\begin{eqnarray}
\vp^\n &=& \mn \left( \vv_\n + \eps_\n \, \vDv\right)\,,
\label{eq:vpn}\\
\vp^\p &=& \mn \left( \vv_\p - \eps_\p \, \vDv \right)\,,
\label{eq:vpp}\\
p_0^\n &=& - \mu^\n + {1\over2}\mn \vv_\n^2 - \vv_\n\cdot\vp^\n\,,\\
p_0^\p &=& - \mu^\p + {1\over2}\mn \vv_\p^2 - \vv_\p\cdot\vp^\p\,,
\end{eqnarray}
where we have defined the dimensionless parameters $\eps_\X$
characterizing entrainment by
\begin{equation}
\eps_\X \equiv {2 \entr\over \mn n_\X}\,.
\end{equation}
Sometimes it is more convenient to use a single entrainment-parameter
$\eps$, which we choose to be $\epsp$, so we have
\begin{equation}
\epsp = \eps\,,\qaq \epsn = {\xp \over 1-\xp}\, \eps\,,
\end{equation}
in terms of the proton fraction $\xp$, which is naturally defined as
\begin{equation}
\xp\equiv {n_\p \over n }\,, \quad\text{with}\quad n\equiv { n_\n + n_\p}\,.
\end{equation}
We note that this definition of the entrainment $\eps$ is different
from another definition, $\epsilon$, which is sometimes found in 
the literature 
(e.g. \citet{lee03:_superfl_r_modes,lindblom00:_r_modes_superfl_neutr_stars}). 
The relation between these two different definitions is simply
(see \citet{prix02:_slow_rot_ns_entrain} for further discussion)
\begin{equation}
\epsilon = {\eps \, n_\p \over n_\n - \eps \, n}\,.
\label{eq:AltEntr}
\end{equation}
We assume that the timescale of oscillations is much shorter than that
of $\beta$-reactions. Therefore strict conservation of neutrons
and protons applies, i.e. we have  
\begin{eqnarray}
\partial_t n_\n + \vnabla\cdot(n_\n \vv_\n ) &=& 0\,, 
\label{eq:Consn}\\ 
\partial_t n_\p + \vnabla\cdot(n_\p \vv_\p ) &=& 0\,.
\label{eq:Consp}
\end{eqnarray}
As shown by \citet{prix02:_variat_I}, 
the equations of motion for the two fluids can be derived from the
Lagrangian density (\ref{eq:DefLagr}) using a ``convective''
variational principle. They can be written in the form
\begin{equation}
\left( \partial_t + \vv_\X\cdot\vnabla \right) \, \vp^\X 
+ p^\X_i \,\vnabla v_\X^i 
- \vnabla Q^\X = {\vf^\X \over n_\X} \,,
\label{eq:EOM0}
\end{equation}
where $\vf^\X$ is the ``external'' force density acting on the fluid
$\X$, and the scalars $Q^X$ are defined as 
\begin{eqnarray}
Q^\X &\equiv& p_0^\X - m^\X \Phi + \vv_\X \cdot \vp^\X \nonumber\\
&=& - \mu^\X +{1\over2}m_\n \vv_\X^2-m_\n\Phi\,.
\end{eqnarray}
In the absence of ``external'' forces acting on the
whole system, the hydrodynamic force densities $\vf^\X$ in (\ref{eq:EOM0})
have to satisfy \mb{$\vf^\n + \vf^\p = 0$} as a Noether identity of the
variational principle. This still allows one to describe a mutual
force $\vf_\mutual$ acting between the two fluids. It  could be
caused, for example, by collisions of the electrons with the neutron vortices
(e.g. cf. \cite{alpar84:_rapid_postglitch}). Such a model would be
characterized by  \mb{$\vf^\n = -\vf^\p = \vf_\mutual$}.
As a first step, however, we only consider the ``free'' limit
and postpone the inclusion of mutual friction and viscosity to future
work. Our ``free'' model is therefore characterized by
\begin{equation}
\vf^\X = 0\,,\quad\textrm{for}\quad \X=\n,\,\p\,.
\end{equation}

\section{The stationary background}

We assume the background to be stationary and axisymmetric, with both
fluids rotating around the $z$-axis with rotation rates $\Om_\n$ and
$\Om_\p$ respectively. Hence
\begin{equation}
\vv_\X = \Om_\X \, \vph\,,\quad \textrm{and}\quad
\vDv = (\Om_\p - \Om_\n)\, \vph\,,
\end{equation}
where $\vph$ is the axial Killing vector, given by
\begin{equation}
\vph = {\partial x^i \over \partial \ph} \, \partial_i = \partial_\ph\,.
\end{equation}
In spherical coordinates, i.e. \mb{$x^i \in \{r,\,\th,\,\ph\}$}, this vector has
the components $\vph^i=(0,\,0,\,1)$, and its norm is  
\mb{$\ph^i\ph_i=r^2\sin^2\th$}.
With the entrainment relations (\ref{eq:vpn}) and (\ref{eq:vpp}) we
can now write the background fluid momenta as
\begin{eqnarray}
\vp^\n &=& \mn \left(\Om_\n - \eps_\n (\Om_\n-\Om_\p) \right)\, \vph\,,\\
\vp^\p &=& \mn \left(\Om_\p - \eps_\p (\Om_\p-\Om_\n) \right)\, \vph\,.
\end{eqnarray}
In the following it will be convenient to introduce as a shorthand notation
the tilde-operator acting on a constituent quantity, $\Om_\X$ say, as follows
\begin{equation}
\til{\Om}_\X \equiv \Om_\X - \eps_\X (\Om_\X -
\Om_\Y)\,,\quad\text{where}\quad \Y \not= \X\,.
\label{eq:DefTilde}
\end{equation}
This allows us to rewrite the background-momenta as
\begin{equation}
\vp^\X = \mn \, \til{\Om}_\X \, \vph\,.
\end{equation}
We restrict our attention to models with uniform rotation, i.e.
\mb{$\vnabla\Om_\X = 0$}, therefore the background vorticities are
\begin{equation}
\vnabla \times \vp^\X = 2 \mn \til{\Om}_\X \, \vz + (\Om_\X - \Om_\Y)\, \vph \times
\vnabla\eps_\X\,, 
\end{equation}
where $\vz$ is the unit vector along the $z$-axis. We see that in the
general case of a varying entrainment $\eps_\X$ and different
background rotation rates \mb{$\Om_\n\not=\Om_\p$}, the vorticities are no
longer aligned with the rotation axis. In other words, they acquire a non-zero
$\theta$-component and the system is in a state which resembles differential rotation.  

As a first step towards a complete understanding of the dynamics of 
rotating two-fluid systems, we will focus on one of the simplest 
possibilities. We make the assumption that the entrainment
$\eps_\X$ is constant throughout the star, which means that we have
\mb{$\vnabla \eps_\X = 0$} and therefore also \mb{$\vnabla\til{\Om}_\X=0$}. 
We note that assuming \emph{both} entrainment parameters $\eps_\X$ to
be constant also requires a constant proton fraction $\xp$.
In the following we therefore consider a \emph{non-stratified} neutron
star model (i.e. \mb{$\vnabla\xp=0$}) with constant entrainment. 
This model is admittedly simplistic, but as we will see in the following, 
it nevertheless allows for a rich phenomenology. 

\section{Linear oscillations}

\subsection{Oscillation equations in harmonic basis}

Assuming uniform rotation and a constant entrainment model as
discussed above, the linear perturbation of the equations of motion
(\ref{eq:EOM0}) can be obtained in the form 
\begin{eqnarray}
\left( \partial_t + \Om_\X \Lie_{\vph} \right) \, {\d\vp^\X  \over \mn}
&+& 2\til{\Om}_\X \vCor_\X + \vnabla \psi_\X = 0\,,
\label{eq:dEuler}
\end{eqnarray}
where the Lie derivative explicitly gives
\mb{$\Lie_{\vph} \,\d p_i = \ph^j\nabla_j \d p_i + \d p_j \nabla_i
  \ph^j$}, and $\d$ represents an Eulerian perturbation. We
have also defined the Coriolis-term $\vCor_\X$ for each of the two fluids as 
\begin{equation}
\vCor_\X \equiv \vz \times \d\vv_\X \,,
\label{eq:DefCoriolis}
\end{equation}
and a scalar potential $\psi_\X$ representing the ``effective'' pressure
perturbation, namely
\begin{equation}
\mn \psi_\X \equiv \d\mu^\X + \mn \,\d\phi + (\vp^\X-\mn\vv_\X)\cdot\d\vv_\X \,.
\end{equation}
The background is assumed to be stationary and axisymmetric, so we can
look for eigenmode solutions of the form  $e^{i(\om t+m\ph)}$, and
the comoving time derivative in (\ref{eq:dEuler}) can  be
directly replaced by   
\begin{equation}
\left(\partial_t + \Om_\X \Lie_{\vph}\right) \rightarrow 
i\left(\om + m \Om_\X \right).
\end{equation}
The practical advantage of using the Lie-derivative $\Lie_{\vph}$ here
is that the substitution \mb{$\Lie_{\vph} \rightarrow i m$} holds for any
geometric object (e.g. a vector as in Eq.~(\ref{eq:dEuler})) with a
$\ph$-dependence of the form $e^{i m \ph}$, while this is only true for the simple
directional derivative $\vph\cdot\vnabla$ when it is
applied to scalars.  

Linear perturbation of the conservation equations (\ref{eq:Consn})
and (\ref{eq:Consp}) leads to
\begin{equation}
\partial_t \d n_\X + \vnabla\cdot\left( n_\X \d\vv_\X + \d n_\X \vv_\X \right) = 0\,.
\label{eq:dConsX}
\end{equation}
In the present analysis we are only interested in inertial modes, which
are characterized by frequencies of the order of the
rotation rate $\Om$. Since this frequency is usually much lower than 
that of the lowest
order p-mode frequency $\om_p$, we can simplify the problem
by using the anelastic  approximation (which effectively  ``filters
out'' the p-modes). As discussed in more detail in Appendix~A, the
anelastic approximation consists of replacing the conservation
equations by    
\begin{equation}
\vnabla\cdot \left(n_\X \, \d\vv_\X \right) = 0 + \O\left( {\om^2
    \over \om_p^2}\right)\,.
\label{eq:dConsXanelastic}
\end{equation}
The lowest-order p-mode frequency $\om_p$ is of the order
of the sound-crossing frequency, i.e. \mb{$\om_p=\O(\co/R)$}, where
$\co$ is an averaged sound speed and $R$ is the stellar radius. For
low-frequency modes such as the inertial modes we can therefore drop
the higher order corrections in (\ref{eq:dConsXanelastic}), which
account for the ``elasticity'' (i.e. compressibility) of matter.
In the following we also restrict ourselves to slowly rotating
backgrounds. Since
\begin{equation}
n_\X = n_\X(r) + \O\left({\Om^2\over 4\pi G \rho_0}\right)\,,
\end{equation}
the star remains spherical if we neglect the centrifugal deformation.
It is important to note the difference between the slow-rotation
approximation, which compares the rotation rate $\Om$ to the
Kepler limit \mb{$\Om_K = \O(\sqrt{4\pi G \rho_0})$}, and the
anelastic approximation, which is relevant for mode frequencies that
are small compared to $\om_p$. 

In order to solve the perturbation equations (\ref{eq:dEuler}) and
(\ref{eq:dConsXanelastic}) for inertial modes, we first transform them
into a pseudo one dimensional problem by expressing all angular dependencies in
terms of the the spherical harmonics $Y_l^m(\th,\ph)$. The spherical
harmonics are the eigenvectors of the angular Laplacian, namely   
\begin{equation}
\nabla^2 Y_l^m(\th,\ph)  = - {l(l+1)\over r^2} Y_l^m(\th,\ph)\,,
\end{equation}
Since these functions form a complete orthonormal basis,  we can expand a 
scalar field $\psi_\X$ as  
\begin{equation}
\psi_\X(r,\th,\ph) = \psi_\X^l(r)\, Y_l^m (\th,\ph)\,,
\label{eq:HarmScalar}
\end{equation}
where  here and in the following automatic summation over repeated
``angular'' indices ($l$, $m$,...) applies. Because of the assumption of  an axisymmetric
background, the various $m$-contributions can be decoupled, and therefore
we can consider each value of $m$ separately. 
In order to express a vector field in a similar manner, we use the 
``harmonic basis'' \mb{$\{\vR_l^m$,\,$\vS_l^m$,\,$\vT_l^m\}$}, which
is defined in terms of the spherical harmonics as
\begin{equation}
\vR_l^m \equiv Y_l^m \, \vnabla r\,,\quad
\vS_l^m \equiv Y_l^m\,,\quad
\vT_l^m \equiv \vS_l^m \times \vnabla r\,.
\label{eq:DefHarmBasis}
\end{equation}
We can then expand the velocity perturbations $\d\vv_\X$ as
\begin{equation}
\d\vv_\X = {W_\X^l(r) \over r}\, \vR_l^m + V_\X^l(r)\, \vS_l^m  
- i U_\X^l \,  \vT_l^m \,.
\label{eq:HarmVector}
\end{equation}
Using the entrainment relations (\ref{eq:vpn}) and (\ref{eq:vpp}), and assuming
a constant entrainment model we obtain
\begin{equation}
{\d\vp^\X \over \mn} = \d\vv_\X - \eps_\X \, \left(\d\vv_\X - \d\vv_\Y \right)  
= \d\til{\vv}_\X\,,
\end{equation}
where we have used the definition (\ref{eq:DefTilde}) of the
tilde-operator. This can  conveniently be
written in the harmonic basis as 
\begin{equation}
{\d\vp^\X \over \mn} = {\til{W}_\X^l \over r}\, \vR_l^m + \til{V}_\X^l \,
\vS_l^m  - i \til{U}_\X^l \, \vT_l^m \,.
\end{equation}
The gradient of a scalar field (\ref{eq:HarmScalar}) is readily expressed as
\begin{equation}
\vnabla \psi_\X = {\psi_\X^{\,l}}'(r) \,\vR_l^m + \psi_\X^{\,l}(r) \,\vS_l^m\,.
\end{equation}
where the prime represents a radial derivative.
The expression for the Coriolis-terms (\ref{eq:DefCoriolis}) in the
harmonic basis is found after a straightforward, but somewhat
laborious, calculation to be 
\begin{eqnarray}
\vCor_\X &=& - {i\over r}\left( m V_\X^l + \beta_k^l U_\X^k \right)\, \vR_l^m \nonumber\\
&& - {i\over l (l+1)} \left(m W_\X^l + m V_\X^l + \gamma_k^l U_\X^k \right)\, \vS_l^m \nonumber\\
&& - {1 \over l (l+1)} \left(\beta_l^k W_\X^k + m U_\X^l + \gamma_k^l   V_\X^k  \right)\,\vT_l^m\,,
\end{eqnarray}
where we sum over the repeated ``angular'' indices ($k$ and $l$), and the constant
matrices $\beta_l^k$ and $\gamma_k^l$ are defined as
\begin{eqnarray}
\beta_l^k &\equiv& l Q_{l+1}\, \d_{k,\,l+1} - (l+1)Q_l \, \d_{k,\,l-1}\,,\\
\gamma_k^l &\equiv& (l^2 -1) Q_l \, \d_{k,\,l-1} + l(l+2) Q_{l+1}\,\d_{k,\,l+1}\,,
\end{eqnarray}
with the usual definition
\begin{equation}
Q_l \equiv \sqrt{ l^2 - m^2 \over 4 \,l^2 - 1}\,.
\end{equation}

\subsection{The general eigenmode equations}

Putting all the pieces together, we can now express 
the complete system of equations (\ref{eq:dEuler}) and (\ref{eq:dConsXanelastic}) 
 in the harmonic basis as
\begin{eqnarray}
r {W_\X^l}' + \left( 1 + r {n_\X' \over n_\X} \right)\, W_\X^l - l (l+1) V_\X^l &=& 0\,,
\label{eq:final1}\\
\kap_\X \til{W}_\X^l - 2 (m V_\X^l + \beta_k^l U_\X^k  ) &=& 2r \,\psih_\X^{l\,'} \,,\\
\kap_\X \til{V}_\X^l - {2 \over l (l+1)} (m W_\X^l + m V_\X^l + \gamma_k^l U_\X^k)
&=& 2 \, \psih_\X^l\,, \\
\kap_\X \til{U}_\X^l - {2\over l (l+1)} (\beta_l^k W_\X^k + m U_\X^l +
\gamma_k^l V_\X^k ) &=& 0\,, 
\label{eq:final4}
\end{eqnarray}
where we defined
\begin{equation}
\kap_\X \equiv {\om + m\Om_\X \over \til{\Om}_\X }\,, \qaq
\psih_\X^l \equiv {i \over 2 \til{\Om}_\X}\, \psi_\X^l\,.
\label{eq:DefKapX}
\end{equation}
We note that this definition of the dimensionless frequencies
$\kap_\X$ reduces to the usual single-fluid definition \mb{$\kap=(\om
  + m\Om)/\Om$} 
in the case of comoving fluids,  or in the absence of entrainment.  In
both of these cases we have \mb{$\til{\Om}_\X\rightarrow\Om_\X$} as
seen from the definition (\ref{eq:DefTilde}).  

The boundary conditions at the centre of the star ($r=0$) consist of the
regularity requirement of the harmonic expansion (\ref{eq:HarmScalar})
and (\ref{eq:HarmVector}), which implies the asymptotic conditions
\begin{equation}
W_\X^l \sim W_\X^l \sim U_\X^l \sim \psi_\X \sim
\O(r^l)\quad\text{as}\quad r \rightarrow 0 \,.
\end{equation}
At the surface ($r=R$) we require another regularity condition
due to the divergent term $n_\X' / n_\X$ in the conservation equations
(\ref{eq:final1}).  As discussed in the Appendix~\ref{sec:Anelastic}, 
this is a consequence of the anelastic
approximation. The
resulting surface boundary condition is therefore
\begin{equation}
W_\X (R) = 0\,,
\end{equation}
i.e. the radial displacement vanishes at the surface.
For models with a vanishing surface density we do not need to impose
an explicit condition on the pressure variation at the surface, as we
have 
\begin{equation}
\Delta P = \d P + \vxi \cdot \vnabla P = \d P = \csum \n_\X\,\d\mu^\X\,.
\end{equation}
 The vanishing of the
Lagrangian pressure perturbation $\Delta P$ is therefore 
ensured provided that the $\d\mu^\X$ are regular at the surface. As our
numerical scheme can only find such regular solutions, this boundary
condition  is implicitly guaranteed to hold.

\subsection{Special case: zero entrainment}
\label{sec:ZeroEntrainment}

We note that the \emph{only} coupling between the neutrons (\mb{$\X=\n$})
and the protons (\mb{$\X=\p$}) in the eigenvalue system
(\ref{eq:final1})--(\ref{eq:final4}) is caused by the entrainment,
cf. the definition of the tilde-operator (\ref{eq:DefTilde}). In
the case of zero entrainment, i.e. $\epsn = \epsp = 0$,  
we obtain two uncoupled eigenvalue systems. Both systems 
are formally identical, and therefore both have the same solutions for
$\kap_\X$, i.e. 
\begin{equation}
\kap_\n = \kap_\p = \kap_\ord\,,
\end{equation}
where $\kap_\ord$ represents the single-fluid solutions.
However, from the definition (\ref{eq:DefKapX}) we see that these
correspond to \emph{different} mode frequencies when the rotation
rates of the two fluids are different, i.e.
\begin{equation}
\om_\n = \left(\kap_\ord  - m\right)\,\Om_\n \,,\qaq
\om_\p = \left(\kap_\ord  - m \right)\,\Om_\p \,,
\end{equation}
which implies that these two solutions cannot form a single mode
solution when \mb{$\Om_\n \not=\Om_\p$}. The two modes in this case therefore
correspond to only one of the two fluids oscillating 
while the other fluid is at rest, i.e. 
\begin{eqnarray}
\om = \om_\n &:& \d\vv_n \not= 0\,,\qaq \d\vv_\p = 0\,,\\
\om = \om_\p &:& \d\vv_n = 0\,,\qaq \d\vv_\p \not= 0\,.
\end{eqnarray}
From the fact that one of the two fluid amplitudes necessarily
vanishes when  \mb{$\eps\rightarrow0$} we deduce that the
corresponding amplitude will actually change sign at this point.  
We can therefore conjecture that if the two fluids were predominantly
in phase before crossing  $\eps=0$, then they will be predominantly in
counter-phase afterwards and vice-versa. We will see in
Sect.~\ref{sec:NumericalResults} that our numerical results 
agree perfectly with this conjecture.

\subsection{The r-mode sub-class}

The subclass of purely axial inertial modes, commonly referred to as
r-modes, has generated a lot of interest due to its strong instability
with respect to gravitational waves
\citep{andersson98:_r_modes,friedman98:_axial_instab}. 
Therefore it is interesting to see if this subclass still exists in
the superfluid case, and how its properties are modified.
A purely axial velocity perturbation is proportional to $\vT_l^m$, so
we set $W_\X^l = V_\X^l = 0$. In this case the equations
(\ref{eq:final1})--(\ref{eq:final4}) reduce to 
\begin{eqnarray}
2\til{\Om}_\X\, \beta_k^l U_\X^k + i r \psi_\X^{l\,'} &=& 0\,, 
\label{eq:axial1}\\
2\til{\Om}_\X \, \gamma_k^l U_\X^k + i l (l+1) \,\psi_\X^l &=& 0\,,
\label{eq:axial2}\\
{1\over 2} \kap_\X l (l+1) \,\til{U}_\X^l -  m \, U_\X^l &=& 0\,.
\label{eq:axial3}
\end{eqnarray}
While the first two equations (\ref{eq:axial1}) and (\ref{eq:axial2})
allow one to calculate the eigenfunctions $U_\X^l$ and $\psi_\X^l$,
they do not constrain the eigenvalue in any way.
The third equation (\ref{eq:axial3}), however, leads to an algebraic
constraint for the existence of a non-trivial axial solution. 
In order to find this constraint, we use the explicit expressions for (\ref{eq:axial3})
for the two fluids, i.e.
\begin{eqnarray}
{1\over 2} l (l+1) \, \kap_\n \left[(1-\eps_\n)\, U^l_\n + \eps_\n U_\p^l \right] - m U_\n^l &=& 0\,,\\
{1\over 2} l (l+1) \, \kap_\p \left[(1-\eps_\p)\, U^l_\p  + \eps_\p U_\n^l \right] - m U_\p^l &=& 0\,,
\end{eqnarray}
from which we can eliminate the $U_\X^l$ (assumed nonzero) to obtain
the following dispersion relation for superfluid r-modes 
\begin{eqnarray}
& &\hspace*{-1cm}\left[l (l+1) (1-\eps_\n) (\om + m \Om_\n) - 2m \til{\Om}_\n \right] \nonumber\\
& & \hspace*{1cm}\times \left[l (l+1) (1-\eps_\p) (\om + m \Om_\p) - 2m \til{\Om}_\p \right] \nonumber\\
&-& l^2 (l+1)^2 \eps_\n \eps_\p (\om + m \Om_\n) (\om + m \Om_\p) = 0\,.
\label{eq:rModeDispersion}
\end{eqnarray}
We note that this corrects the dispersion relation that was 
used by \cite{acp03:_twostream_prl} in a discussion of the superfluid
two-stream instability of the r-modes.

\section{The co-rotating case $\Om_\n = \Om_\p$}
\label{sec:corot}

Before turning to the numerical solution of the general case with
\mb{$\Om_\n\not=\Om_\p$}, it is instructive to study  the special
case of two co-rotating fluids, where we have \mb{$\til{\Om}_\X=
  \Om$}. The linearized perturbation equations (\ref{eq:dEuler})  
and (\ref{eq:dConsXanelastic}) then take the form 
\begin{eqnarray}
i \kap \, \left[\d\vv_\X - \eps_\X (\d\vv_\X - \d\vv_\Y)  \right]
+ 2 \,\vCor_\X + {\vnabla \psi_\X \over \Om}&=& 0\,,\\
\vnabla \cdot \left( n_\X \, \d\vv_\X \right) &=& 0\,,
\end{eqnarray}
where we have defined 
\begin{equation}
\kap \equiv {\om + m\Om \over \Om}\,,
\label{eq:FreqRelation}
\end{equation}
which is the usual dimensionless frequency of inertial modes in the
co-rotating frame. It is interesting to see under which conditions this
system can be separated into purely co- and counter-moving modes. We
therefore introduce the usual variables corresponding to these two
mode-classes, namely 
\begin{equation}
  \begin{array}{r l r l}
\d\vDv \equiv & \d\vv_\p - \d\vv_\n \,,\quad & \d\beta \equiv & \psi_\p - \psi_\n \,,\\
\d\vv \equiv & \xp\,\d\vv_\p + \xn \,\d\vv_\n \,,\quad&
\d\mu \equiv & \xp\,\psi_\p + \xn \,\psi_\n \,.\\
\end{array}
\label{eq:TraditionalCoordinates}
\end{equation}
In terms of these variables the oscillation equations can  be
rewritten as  
\begin{eqnarray}
\vnabla \cdot \left(n \, \d\vv \right) &=& 0\,,\\
i \kap \, \d\vv + 2\vz \times \d\vv + \vnabla \psi &=& \d\beta \,\vnabla \xp\,,\\
\vnabla\cdot\left( n \xn \xp \, \d\vDv  \right) &=& - n \, \d\vv\cdot \vnabla \xp \,,\\
i \gamma^{-1}\,\kap \,\d\vDv + 2 \vz \times \d\vDv + \vnabla \d\beta &=& 0\,,
\end{eqnarray}
where we have defined
\begin{equation}
\gamma \equiv {1\over 1 - \epsn - \epsp} 
= \left( 1- {\eps\over 1-\xp}  \right)^{-1}\,.
\label{eq:DefGamma}
\end{equation}
We see that the variables \mb{$\{\d\vv, \d\mu\}$}, which are
characteristic of ``ordinary''-type modes, decouple from the
``superfluid'' variables \mb{$\{\d\vDv,
  \d\beta\}$},  
if and only if the background model is not stratified, 
i.e. if $\vnabla \xp= 0$. This is exactly the same condition that was found
in the case of a static superfluid neutron star \citep{prix02:_adiab}.  
We further see that in the non-stratified case the equations
governing the two mode-families are equivalent, and the
``ordinary''-type mode frequencies $\kap_\ord$ are therefore related
to the ``superfluid''-type ones by
\begin{equation}
\kap_\sf = \gamma \,\kap_\ord \,,
\label{eq:sfFreq}
\end{equation}
It is well known \citep{bryan89} that the inertial mode frequencies
$\kap\,\Om$ of  an incompressible fluid are bounded (and form a
dense set) in the interval \mb{$[-2\Om,\,2\Om]$}. In the compressible case we
still expect this to hold approximately. This will therefore also be true for the
``ordinary''-type modes in the co-rotating case, but relation
(\ref{eq:sfFreq}) shows that the corresponding interval for the
``superfluid''-type modes will be governed by the factor $\gamma$.
This scale factor  depends only on the proton fraction $\xp$
and the entrainment $\eps$, and can in principle take any value
between \mb{$[-\infty, \,+\infty]$}. For \mb{$\eps<0$} we have
\mb{$\gam\in(0,1)$}, i.e. the ``superfluid''-type mode frequencies
lie closer to the origin than their ``ordinary''-type
counterpart, and they are bounded by a \emph{smaller} interval than
the ``ordinary'' modes. For \mb{$\eps>0$} on the other hand, the
``superfluid''-type mode frequencies lie further away from the
origin than their ``ordinary'' counterparts and their bounding
interval is larger.
We also note that the ``ordinary''-type modes are independent of
the entrainment, as expected from their strictly co-moving 
character. 
If we express the scale factor (\ref{eq:DefGamma}) in terms
of the alternative entrainment parameter $\epsilon$ as defined in
(\ref{eq:AltEntr}), we find
\begin{equation}
\gamma  = 1 + {\epsilon \over \xp}\,.
\end{equation}
Therefore we see that the ``superfluid''-type mode frequencies are
\emph{linear} in $\epsilon$. This has been found previously for the r-mode
subclass by \citet{andersson01:_dyn_superfl_ns}. It was also
observed numerically for inertial modes by
\citet{lee03:_superfl_r_modes} and
\citet{yoshida03:_sf_inertial_modes}, although only as an ``almost'' 
linear dependence. This slight discrepancy is not surprising as their
background model is stratified (i.e. $\xp$ is not constant) and the
above decoupling of the mode-families is therefore expected to hold
only approximately.

\subsection{The r-modes of the co-rotating model}
In the case of a co-rotating background, the r-mode dispersion relation
(\ref{eq:rModeDispersion}) reduces to 
\begin{equation}
\left[(1-\eps_\n) \,\kap - \kap_\ord \right] \, 
\left[(1-\eps_\p) \,\kap - \kap_\ord \right] \, 
- \eps_\n \eps_\p \,\kap^2 = 0\,,
\end{equation}
where 
\begin{equation}
\kap_\ord \equiv {2m \over l (l+1)}\,,
\end{equation}
is the standard single-fluid r-mode frequency. Solving the
quadratic dispersion relation we find that the two r-mode frequencies of
the superfluid problem are
\begin{equation}
\kap = \left\{\kap_\ord, \;\gamma\,\kap_\ord \right\}\,,
\label{eq:cofreqs}
\end{equation}
in agreement with the general relation (\ref{eq:sfFreq}).

In order to find the corresponding eigenfunctions, we eliminate $\psi_\X^l$ from
(\ref{eq:axial1}), (\ref{eq:axial2}) to obtain
\begin{eqnarray}
&&{l-1 \over l} Q_l^m \left[U_\X^{l-1\,'} - l {U_\X^{l-1} \over r}\right] \nonumber \\
&& \hspace*{1cm}+ {l+2 \over l+1} Q_{l+1}^m \left[ U_\X^{l+1\,'} + 
(l+1) { U_\X^{l+1} \over r}\right] = 0\,.
\end{eqnarray}
This equation has to hold for every \mb{$l\ge m$}, and we can therefore
extract the two simultaneous conditions 
\begin{eqnarray}
Q_{l+1}^m \left[ U_\X^{l\,'} - (l+1) \, {U_\X^l \over r}\right] &=& 0\,,\\
Q_l^m \left[U_\X^{l\,'} + l \, {U_\X^l \over r}  \right] &=& 0\,,
\end{eqnarray}
for which the only non-trivial and non-singular solution is
\begin{equation}
U_\X^l = C_\X \, r^{l+1}\,,\quad\text{with}\quad m = l\,.
\end{equation}
Substituting this eigenfunction and the eigenvalues (\ref{eq:cofreqs}) back into
(\ref{eq:axial3}), we find the following relation between the two amplitudes
\begin{eqnarray}
\kap = \kap_\ord &:& C_\p = C_\n \,,
\label{eq:rmodeComoving}\\
\kap = \gamma\, \kap_\ord  &:& C_\p = - {n_\n \over n_\p} \,C_\n \,,
\label{eq:rmodeCountermoving}
\end{eqnarray}
which corresponds to purely co-moving and counter-moving r-modes respectively. 

\section{Numerial results for the general case $\Om_\n\not=\Om_\p$}
\label{sec:NumericalResults}

In the following we choose the proton rotation rate $\Om_\p$ as the
``reference'' rotation rate. This choice is motivated by the fact that
the observed rotation of neutron stars (via pulsar emission) is
thought to be related to the charged components (assumed to co-rotate with the crust), 
while the rotation
rate $\Om_\n$ of the superfluid neutrons is not directly
observable. We  define the relative rotation rate $\R$ as 
\begin{equation}
\R \equiv {\Om_\n - \Om_\p \over \Om_\p}\,.
\end{equation}
With these definitions we can write~(\ref{eq:DefTilde}) as
\begin{eqnarray}
\til{\Om}_\n &=& \Om_\p \left[1 + (1-\eps_\n)\, \R \right]\,,\\
\til{\Om}_\p &=& \Om_\p \left[1 + \eps_\p \, \R \right]\,.
\end{eqnarray}
Further introducing 
\begin{equation}
2 \, \gamn \equiv {1\over 1 + (1-\eps_\n) \,\R}\,,\qaq
2 \, \gamp \equiv {1 \over 1 + \eps_\p \, \R }\,,
\label{eq:DefGamX}
\end{equation}
we can express $\kap_\X$ defined in (\ref{eq:DefKapX}) as
\begin{eqnarray}
\kap_\n &=& 2 \gamn \, \kap_0 + 2 m \gamn \, \R\,,\\ 
\kap_\p &=& 2 \gamp \, \kap_0\,,
\end{eqnarray}
where $\kap_0$ is a dimensionless ``reference frequency'' of the mode, which
we define as
\begin{equation}
\kap_0 \equiv {\om + m\, \Om_\p \over \Om_\p}\,,
\end{equation}
in analogy with the usual single-fluid definition.
With these definitions we can write the system of equations
(\ref{eq:final1})--(\ref{eq:final4}) as a one dimensional infinite
eigenvalue problem for $\kap_0$ in the form
\begin{equation}
\sum_{l=|m|}^\infty \widehat{A}_l \Psi^l = \kap_0 \, \sum_{l=|m|}^\infty \widehat{B}_l \Psi^l\,,
\end{equation}
where the $\widehat{A}_l$ and $\widehat{B}_l$ are linear operators and $\Psi^l$ is the eigenvector
\begin{equation}
\Psi^l = \left\{W_\n^l,\, W_\p^l,\, V_\n^l,\, V_\p^l,\, U_\n^l,\, U_\p^l,\, \psih_\n^l,\,
  \psih_\p^l  \right\}\,. 
\end{equation}
The explicit form of these equations is given in
Appendix~\ref{sec:ExplicitEquations}.
By taking the sum over $l$ only up to a finite value $l_\mathrm{max}$, we can solve
the resulting finite eigenvalue problem using the LSB spectral solver,
which is based on the efficient incomplete Arnoldi-Chebychev algorithm. 
This is the same method that \citet{prix02:_adiab} used to study
nonradial oscillations of non-rotating stars.

Most of the numerical calculations in the following have been
performed for both a uniform-density background (i.e. polytropic index
\mb{$N=0$}) and a polytropic background with $N=1$ (for each of the fluids). 
The results are
quite similar and we therefore only present the polytropic case here. 
Furthermore, we only considered the case \mb{$m=2$}, which is expected
to be the most relevant for gravitational-wave emission.
The results for higher values of $m$ are not expected to show
any qualitative differences. 
In all of the following sections except for
Sect.~\ref{sec:twostream},  we use a neutron star model with
``canonical'' values \mb{$\xp=0.1$}  for the proton fraction and
\mb{$\eps=0.6$} for the entrainment, which conveniently results in a scaling-factor
(\ref{eq:DefGamma})  of \mb{$\gamma=3$}. 
In Sect.~\ref{sec:twostream} on the two-stream instability we choose
these values to be \mb{$\xp=0.2$} and \mb{$\eps=-2$}, which leads to 
\mb{$\gamma \approx 0.2857$}. 

For the low-order inertial modes considered in this paper a radial
resolution of 30 Chebychev polynomials and an angular resolution of
about 10 spherical harmonics is used in most cases, which proves
sufficient to obtain a numerical precision of the order of $10^{-6}$.
In the case of r-modes, we compared our numerical results to a
direct evaluation of the dispersion relation
(\ref{eq:rModeDispersion}) and found an agreement better than
$10^{-6}$ in all cases considered. 
In the co-rotating case (\mb{$\R=0$}) our numerical results  for the
``ordinary'' modes agree perfectly (up to the given precision
\mb{$\sim   10^{-6}$}) with the single-fluid results in the
literature \citep[e.g.][]{lockitch99:_r-modes}, and the ``superfluid''
modes satisfy the relation~(\ref{eq:sfFreq}) as expected.

\subsection{Angular convergence and inertial-mode labelling}

It was shown by \citet{lockitch99:_r-modes} that for $m\not=0$ the
lowest non-zero $l$-coefficient in the harmonic expansions
(\ref{eq:HarmScalar}) and (\ref{eq:HarmVector}) is necessarily
$l=|m|$. Furthermore, in 
the case of a uniform density background model (i.e. \mb{$N=0$}), it
is known that the harmonic expansion (\ref{eq:HarmVector}) of the solution
$\d\vv$ stops at a finite $l_0$. In fact, the corresponding coefficients
can be calculated analytically. There are always exactly
\mb{$j\equiv l_0 - |m| + 1$} mode-solutions for any given \mb{$m\not=
  0$} 
and \mb{$l_0\ge |m|$}.  In the case of a polytropic background
with $N=1$  the solution turns out to be quite similar to the uniform case, except that the expansion
does not stop after a finite number of terms. Instead it converges
exponentially beyond \mb{$l = l_0$}. 
\begin{figure}
  \psfrag{lm1}{$l - m + 1$}
  \psfrag{logcl}{$\log_{10}\left|c_l\right|$}
  \includegraphics[width=\hsize,clip]{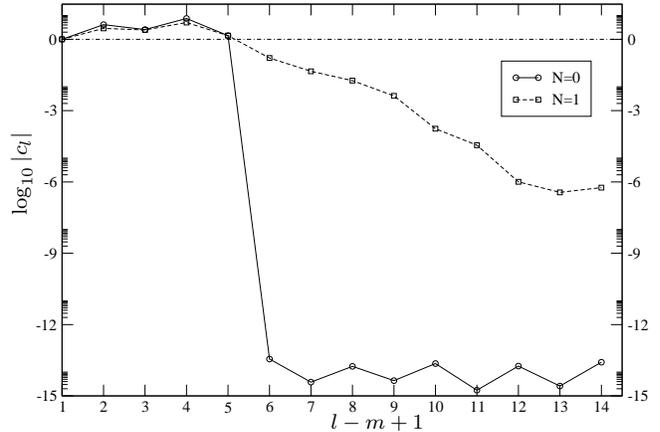}

  \caption{Angular convergence of the (``ordinary'') inertial mode
  with \mb{$m=2$} and \mb{$j\equiv l_0 -m +1 = 5$} (which we label
  $^5_2I^\ord_{(1)}$) at $\R=0$ for the  homogeneous background model
  (\mb{$N=0$}) and the \mb{$N=1$} polytrope. The plotted quantity
  $c_l$ represents the magnitude of the harmonic expansion
  coefficients (\ref{eq:HarmScalar}) and (\ref{eq:HarmVector}) of the eigenmode. }
\label{fig:AngConv}
\end{figure}
This behaviour is illustrated in Fig.~\ref{fig:AngConv}, which shows
the angular expansion coefficients of the (axial-led, $m=2$) inertial
mode with \mb{$\kap= -1.308$} for $N=0$ and the analogous
mode \mb{$\kap = -1.43392$} for $N=1$. 
We see that in the uniform background case there is a sharp drop after
$j = 5$, as the higher-order coefficients are analytically zero, while
in the polytropic case we observe an exponential falloff.
The quantitative and qualitative similarity to the uniform model allows
one to associate the modes of the polytropic model with
corresponding modes of the uniform model. In case of doubt it should
always be possible to associate modes via a continuous transformation
of $N$. We can therefore conveniently label the modes  by their
``quantum numbers'' $m$, $j$ and an additional index \mb{$n \in [1,
  j]$} 
accounting for the $j$ different solutions at given $m$ and $j$. As a
convention we choose to order the modes by increasing frequency $\kap$, so
we label the inertial modes as 
\begin{equation}
^{\;j}_m{I}_{(n)} : {_m^{\;j}{I}_{(1)}} <  {_m^{\;j}{I}_{(2)}} < ... <  {_m^{\;j}{I}_{(j)}}\,,
\end{equation}
where the inequalities obviously refer to the eigenfrequency of the
corresponding mode. 

\subsection{The effect of relative rotation $\R$}

\begin{figure}

{
\psfrag{Rel0.0}{$\R=0.0$}
\psfrag{Rel0.1}{$\R=0.1$}
\psfrag{Un3}[][][0.8][0]{$U_\n^3$}
\psfrag{Up3}[][][0.8][0]{$U_\p^3$}
\psfrag{Vn2}[][][0.8][0]{$V_\n^2$}
\psfrag{Vp2}[][][0.8][0]{$V_\p^2$}
\psfrag{Wn2}[][][0.8][0]{$W_\n^2$}
\psfrag{Wp2}[][][0.8][0]{$W_\p^2$}
\includegraphics[width=\hsize,clip]{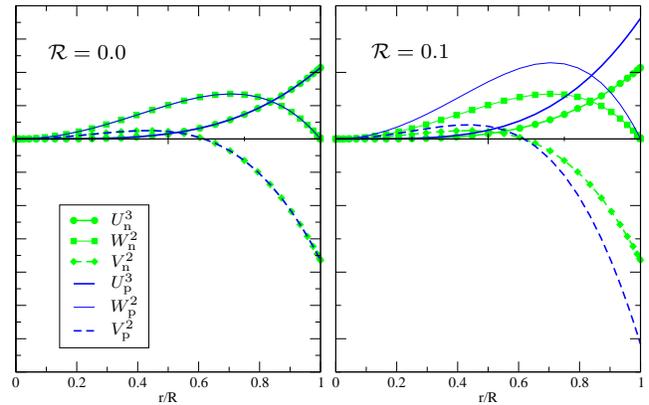}
}
  \caption{The polar-led ``ordinary'' inertial mode $^2_2I_{(1)}^\ord$ for
    $\xp=0.1$ and $\eps=0.6$ for a co-rotating background $\R=0$ (left
    panel) and $\R=0.1$ (right panel). The next higher
    $l$-contributions are one order of magnitude smaller and
    are not included in this graph.}  
\label{fig:I2o1_R}
\end{figure}

We have seen in Sect.~\ref{sec:corot} that, in the co-rotating case the 
inertial modes of non-stratified stars can be separated into purely
co- and counter-moving families. This is no longer true when
we allow for a non-zero relative rotation \mb{$\R\not=0$}. Similar to
stratification (cf. \cite{prix02:_adiab}), the relative rotation
introduces a coupling between these mode-families, leading to a
deviation from the strictly co- and counter-moving nature of the
modes. This is shown in  Fig.~\ref{fig:I2o1_R}. 

\begin{figure}
  \psfrag{Up_R--------------------}{$U_\p:\;\forall\,\R$}
  \psfrag{Un_R0}{$U_\n:\;\R=0$}
  \psfrag{Un_R-0.1}{$U_\n:\;\R=-0.1$}
  \psfrag{Un_R-0.4}{$U_\n:\;\R=-0.4$}
  \psfrag{Un_R-0.6}{$U_\n:\;\R=-0.6$}
\includegraphics[width=\hsize,clip]{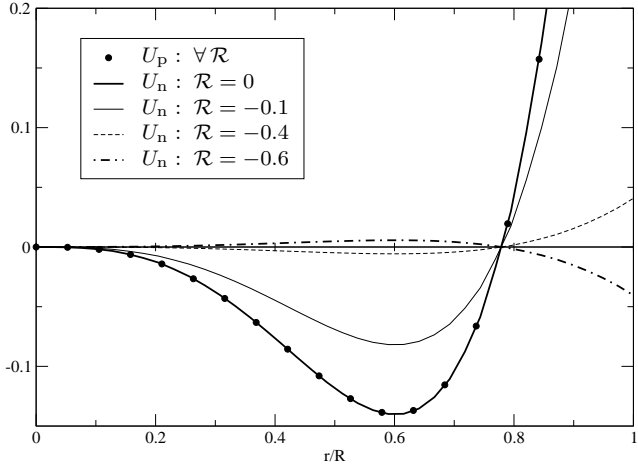}

\caption{Neutron- and proton-amplitudes for different relative
  rotations $\R$. The plot shows the $l=2$ components $U_\n$ and
  $U_\p$ of the mode \mb{$^3_2I_{(2)}^\ord$} for
  \mb{$\R=0,\,-0.1,\,-0.4,\,-0.6$}. The normalization is such that
  \mb{$U_\p(R)=1$}, for which $U_\p$ is seen to be invariant under
  changes of $\R$.}

\label{fig:PhaseChange}
\end{figure}

\begin{figure}
  \includegraphics[width=\hsize,clip]{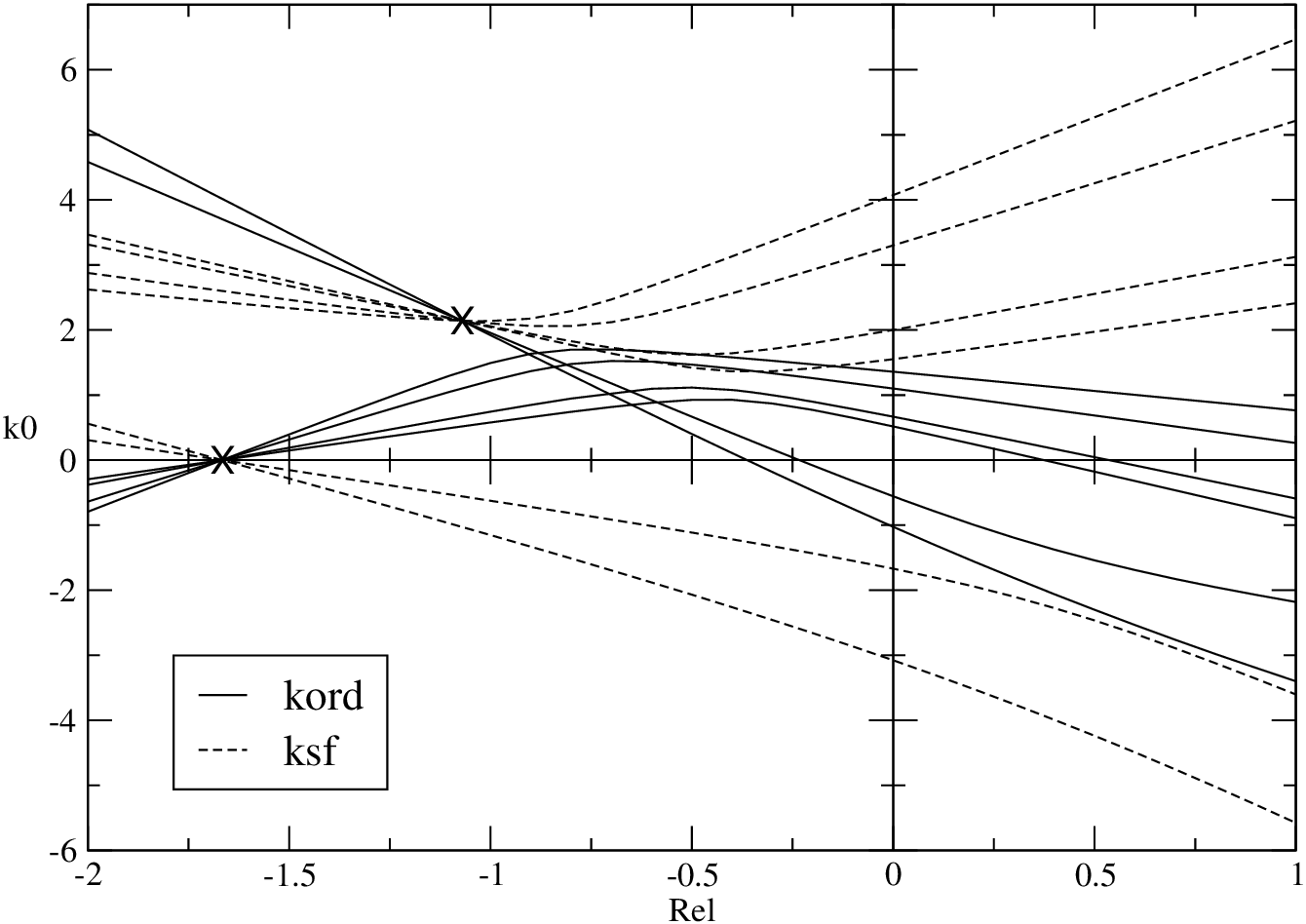}

  \caption{Frequencies $\kap_0$ of ``ordinary'' (--) and ``superfluid'' (-\,-) $m=2$ inertial
    modes as functions of the relative rotation rate $\R$ for an $N=1$ polytropic background model
    with $\xp=0.1$ and $\eps=0.6$. The two common crossing
    points~(\ref{eq:1}) are marked by 'x'. The modes presented
    here are the 6 lowest order inertial modes, $_2^1I_{(1)}$ to $_2^3I_{(3)}$.} 
\label{fig:PolyFreqR}
\end{figure}

Compared to the effect of stratification, however, the mode-coupling
induced by the relative rotation $\R$ (in the absence of
stratification) seems to be of a much weaker nature. Although the two
fluids are no longer strictly co- or counter-moving, they always have
a well-defined phase-relation, in the sense that they are either
strictly in phase or in counter-phase. Changing $\R$ does not change
the position of the nodes of the mode. This can be seen in Fig.~\ref{fig:PhaseChange}
which illustrates the transition of the $^3_2I_{(2)}^\ord$ mode being in phase to
being in counter phase when varying $\R$.
Furthermore, this coupling does not lead to general avoided crossings
between inertial modes, as can be seen in Fig.~\ref{fig:PolyFreqR}
in which we show the mode frequencies of the lowest-order inertial
modes as functions of the relative rotation rate $\R$.

A striking feature of this graph is that there are two common
crossing-points for the mode frequencies. This can be understood as 
follows: the system of equations 
(\ref{eq:LSBfirst})--(\ref{eq:LSBlast}) has a singularity when 
\mb{$\gamX\rightarrow\infty$} for $\X=\n$ or $\X=\p$. We see 
from (\ref{eq:DefGamX}) that this happens at the relative
rotation rates  \mb{$\R = - {1\over 1-\epsn}$} and \mb{$\R=-{1\over
    \epsp}$}, respectively. 
At these singular points the system of equations reduces to the
following constraints: 
\begin{eqnarray}
\gamn\rightarrow \infty &:& \kap_0 = -m \,\R\,,\quad\text{or}\quad
W_\n^l = -{\epsn \over 1-\epsn}\, W_\p^l\,,
\label{eq:CritialPoint1}\\
\gamp \rightarrow \infty &:& \kap_0 = 0\,,\quad\text{or}\quad
W_\p^l = - {\epsp \over 1-\epsp}\, W_\n^l\,,
\label{eq:CritialPoint2}
\end{eqnarray}
and similar amplitude constraints  hold for $V_\X^l$ and $U_\X^l$.
The two common crossing points therefore have to be 
\begin{equation}
  \label{eq:1}
(\R,\,\kap_0) = \{ (-{1\over \epsp},\, 0) \,,\;\;
({1\over 1-\epsn},\, -m\R) \}\,.
\end{equation}
These points are are marked by  'x' in our various  frequency plots.
The solutions at these critical relative rotation rates fall
into two classes: modes  that cross at the common crossing point,
and modes that satisfy the amplitude relations
(\ref{eq:CritialPoint1}) or (\ref{eq:CritialPoint2}).  
These analytical results agree perfectly well with the numerical findings
and provide a good consistency check of our numerical results.  We
note that while in Fig.~\ref{fig:PolyFreqR} it seems as if each of the
modes necessarily passes through one of the two crossing points, this
is not generally the case, as will be seen in Fig.~\ref{fig:InstR_Poly}
for a different choice of parameters.

It is interesting to note that the two critical relative rotation
rates correspond to the vanishing of the angular momentum of one of
the two fluids, i.e. $\gamn\rightarrow\infty$ corresponds to
$\vp^\n=0$ and $\gamp\rightarrow\infty$ is equivalent to $\vp^\p=0$. This is
obviously an effect of the entrainment: the fluid is rotating but has 
zero angular momentum! As a result, the Coriolis force acting on this
fluid vanishes and the mode becomes stationary in the reference frame of the
respective fluid. As we have chosen $\Om_\p$ as our reference 
rotation, we find $\kap_0=0$ for $\vp^\p=0$. The nonzero crossing
frequency \mb{$\kap_0= -m\R$} for $\vp^\n=0$  simply corresponds to the
zero frequency in the neutron-frame observed in the proton-frame.

\begin{figure*}

  \includegraphics[width=\hsize,clip]{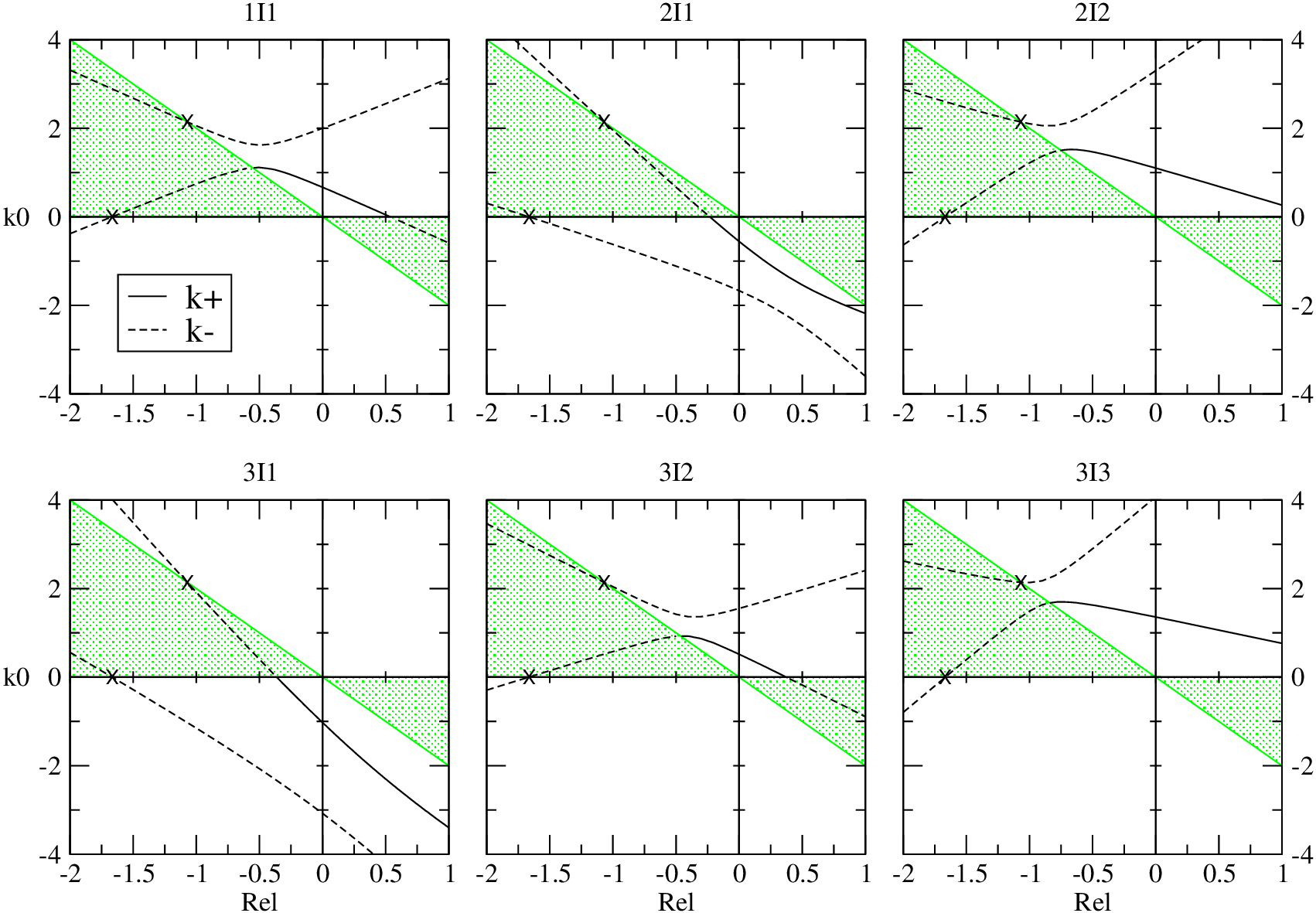}
  \caption{Avoided crossings between ``ordinary'' and ``superfluid''
    (defined at \mb{$\R=0$}) inertial modes as functions of the
    relative rotation rate $\R$. The common
    crossing-points~(\ref{eq:1}) are marked by 'x' and the shaded
    areas corresponds to \mb{$\kap\,(\kap + m \R) < 0$}. The labels
    $\kap_{+}$ and $\kap_{-}$ indicate if the two fluids are in phase
    or in counter-phase respectively.}  

\label{fig:PolyAvoidR}
\end{figure*}

While there are no general avoided crossings, the coupling induced by
$\R$  \emph{does} lead to avoided crossings between corresponding
``mode-pairs'', i.e. between the ``ordinary'' mode and its
``superfluid'' counterpart, as can be seen in
Fig.~\ref{fig:PolyAvoidR}.
We note that the labelling $\kap_\ord$ and $\kap_\sf$ used in
\ref{fig:PolyFreqR} to refer to ``ordinary'' or ``superfluid'' modes
is defined by continuing the mode from
\mb{$\R=0$}. This labelling is somewhat arbitrary, however, as for
\mb{$\R\not=0$} it does not reflect the co- or counter-moving
nature of the mode. Neither does it imply the mode to be in phase or
in counter-phase, as can be seen from Fig.~\ref{fig:PhaseChange}. In
Fig,~\ref{fig:PolyAvoidR} and in the following it will often be more
interesting to indicate the \emph{phase-character} of a mode,
so we will write $\kap_{+}$ for modes with in-phase fluid motion, and
$\kap_{-}$ for modes where the fluids are in counter-phase.

As we have already seen in Fig.~\ref{fig:PhaseChange}, the
relative phase is not an invariant property of the ``ordinary'' or
``superfluid'' mode families. For example, in
Fig.~\ref{fig:PolyAvoidR} the ``superfluid'' modes are always in  
counter-phase, while the ``ordinary'' mode is in phase in a certain
region but in counter-phase in another. We note, however, that the
ordinary mode necessarily has to
be in phase in \mb{$\R=0$}, as we know analytically (see
Sect.~\ref{sec:corot}) that at this point the two mode-families have
strict co- and counter-moving character. 

Let us consider the relation between the pattern-speed 
\mb{$\dot{\ph} = - {\om \over m}$} of the mode and the two rotation
rates $\Om_\n$ and $\Om_\p$. In particular, we are interested in the
region where the pattern-speed of the mode lies in between the
rotation rates of the two fluids, such that it would appear prograde
when viewed in one fluid frame and retrograde in the other. One can see that
this ``mixed'' region is characterized by the condition 
\begin{equation}
    \label{eq:2}
    \kap\, \left(\kap + m \R \right) < 0\,.
  \end{equation}
This ``mixed'' region is indicated in Fig.~\ref{fig:PolyAvoidR} and
Fig.~\ref{fig:InstR_Poly} as a shaded area, and we observe that the
change of the phase-character of modes only happens when the mode
frequency crosses into or out of the ``mixed'' region. There seems to be
no phase-change, however, if the crossing takes place via one of the
two common crossing points (\ref{eq:1}), which are indicated by 
'x' in these figures.  
We can try to understand this as follows:
when a mode crosses into or out of the ``mixed'' region, it means
that its frequency vanishes and changes sign in one of the two fluid
frames. In general the Coriolis force of the corresponding fluid is
nonzero in this point, therefore the frequency can only be zero if the
fluid ceases to move. The corresponding fluid 
eigenfunctions
therefore undergo a sign-change, which results in the phase-change of
the mode. 
In the special case where the crossing happens via one of the two
common crossing points, however, the Coriolis-force \emph{does} vanish
at this point and subsequently changes sign, therefore the mode-amplitude cannot
change sign and the crossing takes place without a phase change.

\subsection{Varying the entrainment}

\begin{figure*}

  \includegraphics[width=\hsize,clip]{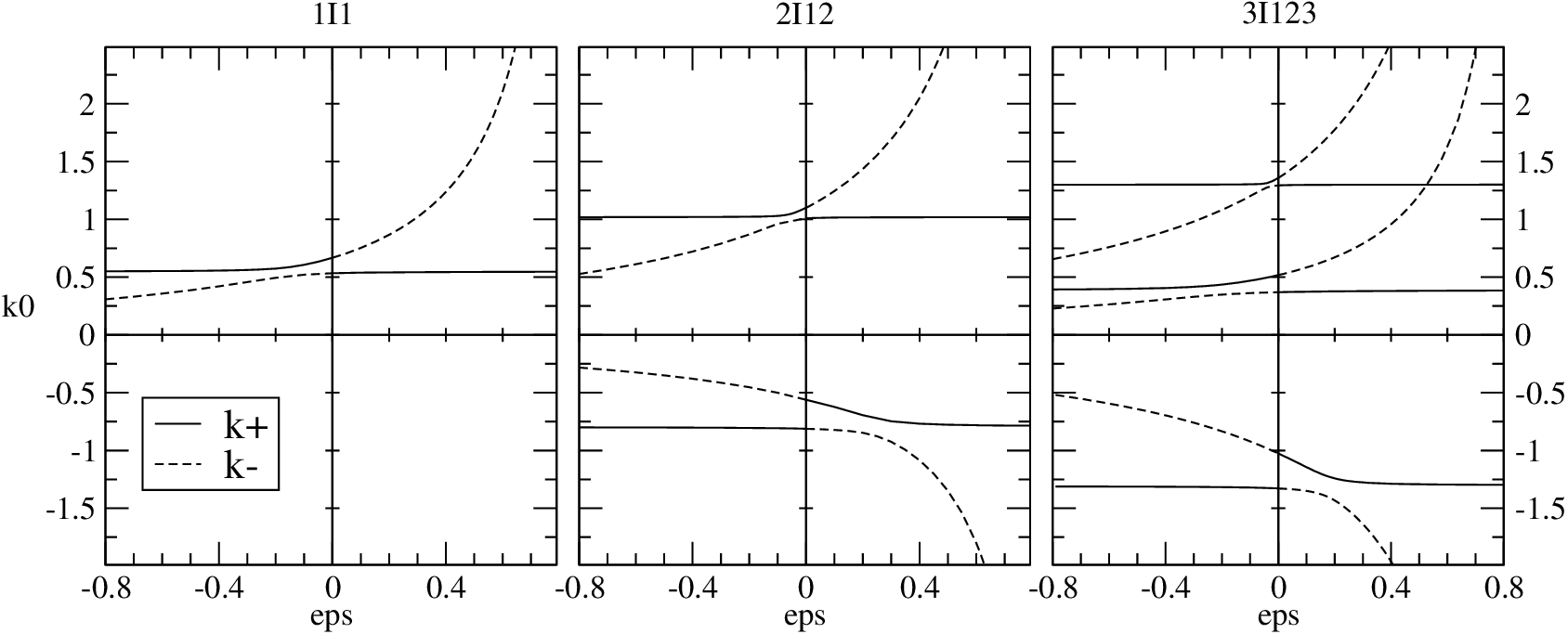}
 
  \caption{Avoided crossings between ``ordinary'' and ``superfluid'' $m=2$ inertial
    modes as functions of entrainment $\eps$ for the $N=1$ polytropic background model and a
    relative rotation rate of $\R=0.1$. The labels $\kap_{+}$ and
    $\kap_{-}$ refer to the phase-character of the modes.} 
\label{fig:PolyAvoidEps}
\end{figure*}

In Fig.~\ref{fig:PolyAvoidEps} we have plotted  the mode-frequencies
as functions of the entrainment $\eps$. Similar to the avoided
crossings as functions of $\R$ shown in Fig.~\ref{fig:PolyAvoidR}, we
observe that there are only ``pairwise'' avoided crossings,
i.e. between an ``ordinary'' and the corresponding ``superfluid'' 
mode.  We further note that the crossing of the zero-entrainment axis
($\eps=0$) is rather special, as can be understood from the
discussion in Sect.~\ref{sec:ZeroEntrainment}.  At $\eps=0$, one of 
the two fluid-amplitudes is necessarily zero, and therefore the crossing
of the $\eps=0$ axis induces a phase-change between the two fluids.
This is exactly the behaviour observed numerically for the modes
shown in Fig.~\ref{fig:PolyAvoidEps}. 

\subsection{The two-stream instability}
\label{sec:twostream}

It was recently discovered \citep{andersson02:_twostream} that  superfluid
systems may, quite generally, suffer a so-called ``two-stream
instability''. In the present context, this instability 
would  set in when the relative velocity between the
two fluids exceeds a certain critical level. This mechanism was
suggested as a possible mechanism for triggering pulsar 
glitches \citep{acp03:_twostream_prl}.
Unfortunately, the dispersion relation for superfluid r-modes on which 
the analysis of \cite{acp03:_twostream_prl} was based is incorrect,
affecting the various estimates for the onset and growth 
of the instability (for a detailed discussion, see \cite{andersson02:_twostream}).

\begin{figure*}

  \includegraphics[width=\hsize,clip]{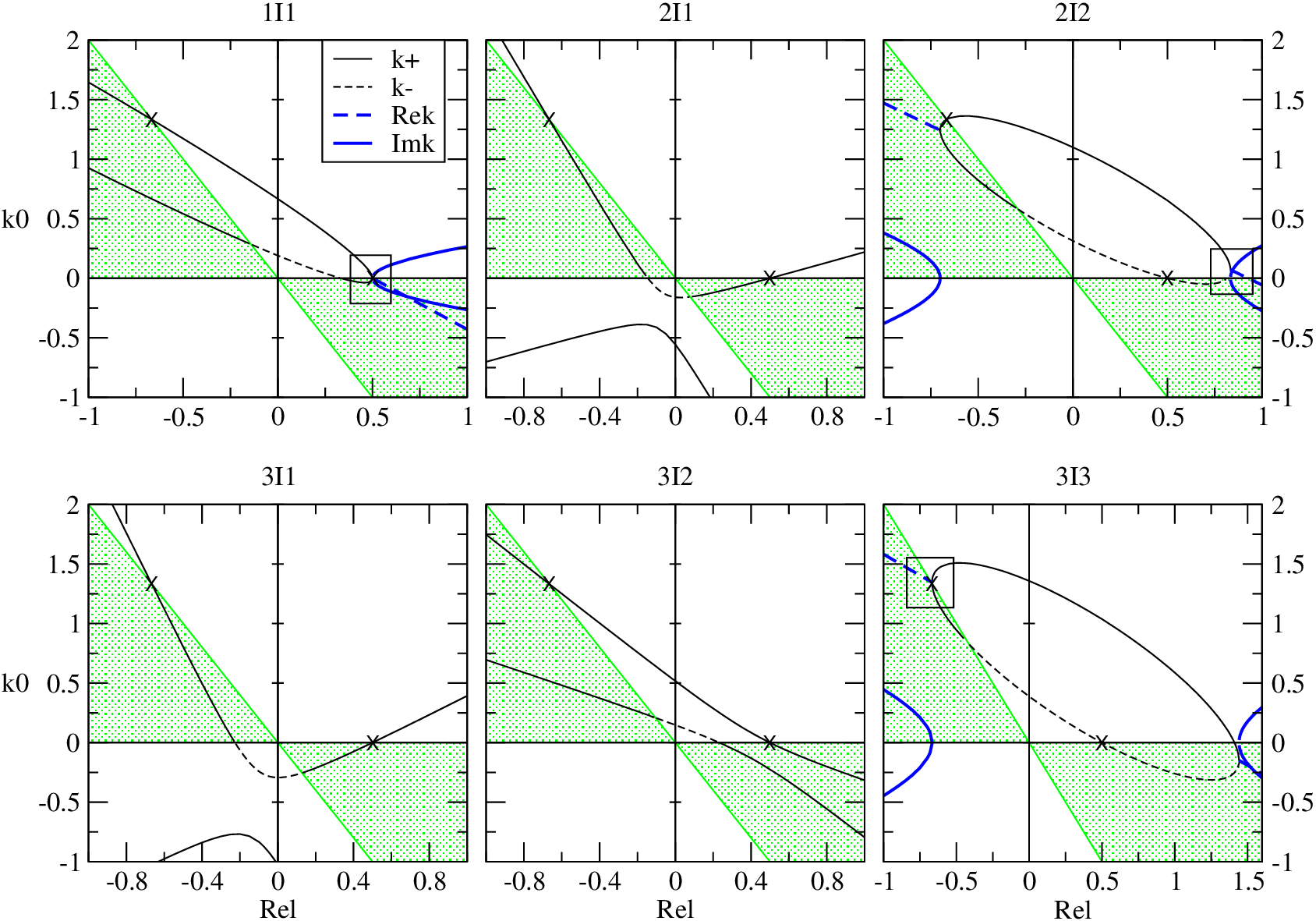}

  \caption{Inertial-mode frequencies as functions of the relative rotation
    $\R$, with \mb{$\xp=0.2$} and $\eps=-2$. The common crossing
    points~(\ref{eq:1}) are marked by 'x'. Merger of two real
    frequencies leads to a complex-conjugate pair, and signals the onset of instability.
    The shaded areas indicate the ``mixed'' regions where
    \mb{$\kap\,(\kap + m\R) < 0$}. The labels $\kap_{+}$ and $\kap_{-}$ indicate
    if the two fluids are in phase or in counter-phase, respectively.
    The boxes indicate regions that we zoom into in Fig.~\ref{fig:InstR_Poly_zoom}.}

  \label{fig:InstR_Poly}
\end{figure*}

As the general instability mechanism discussed in
\cite{andersson02:_twostream} remains sound, we expect to find inertial modes
that become unstable beyond a critical relative rotation rate $\R$. 
For the parameter values chosen for Fig.~\ref{fig:PolyAvoidR}, no
such instabilities were observed within the interval \mb{$-2\leq\R \leq
  1$} that was
considered. However, using the dispersion relation 
(\ref{eq:rModeDispersion}), we can identify a more instability-prone
region to be, for example, a proton fraction of \mb{$\xp=0.2$}
and an entrainment of \mb{$\eps=-2$}. In the neutron-star
core the entrainment $\eps$ is generally expected to be positive, but
a negative entrainment is nevertheless not unphysical. 
Superfluid $^4$He, for example, has negative entrainment, and this is
also expected to be the case for the neutron superfluid in the
neutron-star crust \citep{chamel03:_effec}. While the present example serves
 only as a consistency check and proof of principle, we emphasize
that these parameter-values are not completely unphysical.
In Fig.~\ref{fig:InstR_Poly} we plot the frequencies of the
lowest-order inertial modes as functions of the $\R$ for this choice
of parameters. We see that now
the r-mode $^1_1I_{(1)}$, and the inertial modes $^2_2I_{(2)}$ and
$^3_2I_{(3)}$ do indeed undergo an instability via the merger of the
``ordinary''-type mode with its ``superfluid'' counterpart. After this
merger the two mode-frequencies are complex conjugates, which is 
to be expected from the time-symmetry of the problem. The real part
of $\kap_0$ is strictly linear in $\R$ in the instability region.
\begin{figure*}

  \includegraphics[width=\hsize,clip]{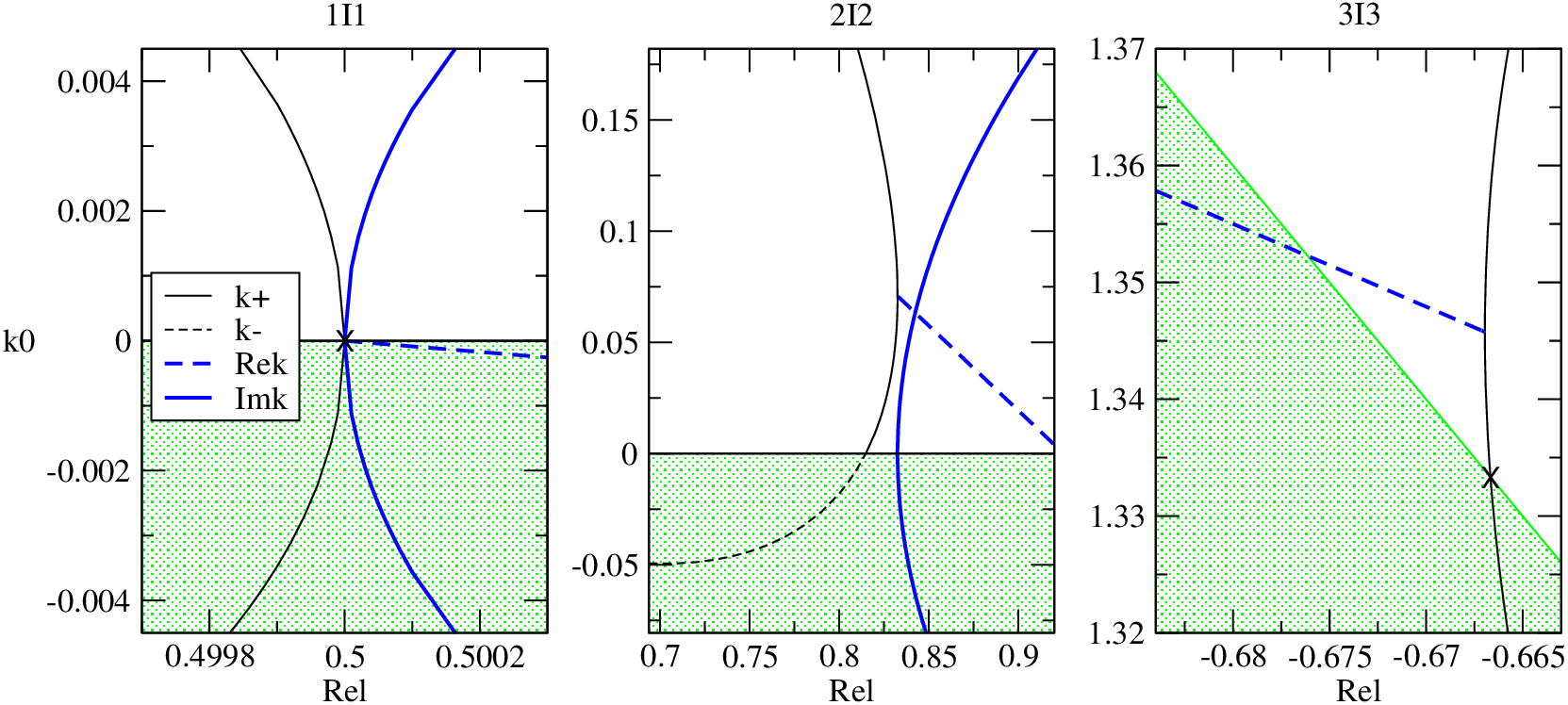}

  \caption{Close up of  the vicinity of the three instability points
    marked by boxes in Fig.~\ref{fig:InstR_Poly}. The common crossing
    points~(\ref{eq:1}) are marked by 'x'. The labels $\kap_{+}$ and
    $\kap_{-}$ indicate if the two fluids are in phase or in
    counter-phase respectively.} 

  \label{fig:InstR_Poly_zoom}
\end{figure*}
For the present set of parameters, the two common
crossing points given by (\ref{eq:1}) are \mb{$(\R,\,\kap_0) = (0.5,
  0)$} and \mb{$(-0.6667, 1.3333)$}. 
These points are marked by 'x' in Fig.~\ref{fig:InstR_Poly}. 
Interestingly, the instability point of the r-mode coincides (up to
numerical precision $\sim 10^{-6}$) with  one of the common crossing points
discussed earlier, namely the one at which the proton-fluid angular
momentum vanishes. Using the analytic r-mode dispersion
relation~(\ref{eq:rModeDispersion}), we can  verify that the
instability occurs \emph{exactly} at the crossing point $(0.5, 0)$.
However, this is clearly seen not to be the case for the higher order
inertial modes. 
One might expect the instabilities to
occur in one of the ``mixed'' regions, as the mode is then
prograde in one fluid frame and retrograde in the other (cf. for
example \citet{pierce74:_almost_all_about_waves}).  
This, however, is not always the case, as illustrated in
Fig.~\ref{fig:InstR_Poly_zoom}. We see that onset of the instabilities
of the $^2_2I_{(2)}$ and  $^3_2I_{(3)}$ modes occur slightly  
outside the ``mixed'' region. Given the numerical precision of
$\le 10^{-6}$, this should not be due to numerical errors.
This observation serves as a strong motivation for a study into the 
stability properties of rotating multi-fluid systems. 
It would be desirable to attempt a derivation of useful 
instability criteria, eg. analogous to those derived by
\cite{friedman78:_secul_instab} for the single fluid problem.  

\section{Discussion}

In this paper we have derived the equations that govern inertial modes
of a slowly rotating superfluid neutron star model in the anelastic
approximation. These equations are more general than ones used in previous
studies since they allow for general non co-rotating
backgrounds \mb{$\Om_\n \not=\Om_\p$}. We have discussed analytically
the special cases of co-rotation and zero entrainment. The obtained
analytical results were then 
confirmed by, and thus served as important benchmark tests for, our numerical 
calculations. We studied numerically the dependence of the mode frequencies
on the relative rotation rate and entrainment, and found 
avoided crossings between mode-pairs. The ``phase character'' of the
modes was found to be rather complex, in the sense that it can change
when crossing into or out of a ``mixed region''. In a ``mixed region''
the mode frequencies lie in between the two background
rotation rates.
We have also confirmed, for the first time in a complete
mode-calculation, the existence of the superfluid two-stream
instability.
We have studied the onset of this instability as a function of
relative rotation rate, and found that contrary to intuitive
expectations, the onset can sometimes take place slightly outside the
``mixed region''. 

The complicated problem of  oscillations of rotating multi-fluid systems
provides many challenges that should inspire  future work.
More detailed models should allow for stratified backgrounds as 
this would be 
closer to a realistic neutron star model. Stratification is expected to 
lead to a substantially more complex character of the mode-spectrum.
In particular, there is likely to  be avoided crossings between \emph{all} modes and
the modes will no longer be of purely ``in-phase'' or in
``counter-phase''-character. This has already been observed in the studies by
\cite{lee03:_superfl_r_modes} and \cite{yoshida03:_sf_inertial_modes}
in the purely co-rotating case.  
It would also be interesting to move beyond both the anelastic
approximation and the slow-rotation approximation, in order to be 
able to consider rapidly spinning stars. One should also account for the presence of 
an elastic crust, perhaps penetrated by a neutron superfluid, and 
include dissipative processes like mutual friction
and beta-reactions between the two fluids.
Another issue that needs to be studied in detail is the potential gravitational-radiation
instability of the various modes, and a suitable adaptation of the CFS
instability criterion 
\citep{chandrasekhar70:_grav_instab,friedman78:_secul_instab} to non
co-rotating backgrounds. This should also help shed light 
on the two-stream instability, the true physical relevance of which is
difficult to assess at the present time. 

\section*{Acknowledgments}
 We wish to thank L.~Valdettaro and M.~Rieutord for allowing us to use
 their LSB code, and we further thank B.~Dintrans and M.~Rieutord for
 valuable discussions and help with the numerics.

 RP and NA acknowledge support from the EU Programme 'Improving the Human 
 Research Potential and the Socio-Economic Knowledge Base' (Research
 Training Network Contract HPRN-CT-2000-00137).
 GLC gratefully acknowledges support from NSF grant PHYS-0140138.
NA is grateful for generous support from a  Philip Leverhulme Prize fellowship.

\bibliography{biblio}

\appendix

\section{The anelastic approximation}
\label{sec:Anelastic} 

The anelastic approximation was first introduced in atmospheric
physics \citep{batchelor53,ogura62} and has since also been widely
used in the study of  stellar oscillations and convection. A more
detailed analysis of the quality and justification of this
approximation in the case of g-modes can be found in
\citet{dintrans01:_anelastic} and \citet{rieutord02:_anelastic}, and
it has also been used recently in the study of inertial modes
\citep{villain02:_inert}.  
The anelastic approximation applies for modes with frequencies which are
small
compared to the inverse of the sound-crossing time of the star, which
characterizes the lowest-order p-mode frequency.
High frequency modes such as p-modes are effectively ``filtered out''
by this approximation, so that only low-frequency modes like inertial
modes or g-modes remain. We will now briefly sketch how this
approximation works in the study of inertial modes of a barotropic
star. We start from the linear perturbation equations for a uniformly
rotating barotrope, assuming an eigenmode solution of the form $e^{i\left( \om \,t + m\, \ph\right)}$,
which yields
\begin{eqnarray}
i \left(\om + m\,\Om \right)\, \d\n + \vnabla\cdot\left( n\, \d\vv \right) &=& 0\,,\\
i \left(\om + m\,\Om \right)\, \d\vv + 2\Om\, \vz\times\d\vv + \vnabla\left( \d\mut +
  \d\phi\right) &=& 0\,,
\end{eqnarray}
We choose an average sound speed $\co$ as the natural velocity scale, and the stellar
radius $R$ as length scale, which implies the sound crossing time $R/\co$ as the natural
time scale. Therefore the dimensionless mode-frequency is
\begin{equation}
\omh \equiv {\om\over \co/R}\,.
\end{equation}
Inertial modes have the property that their frequencies are of the order of $\Om$, so we
introduce
\begin{equation}
\zeta \equiv {\Om\over \om} = \O(1)\,.
\end{equation}
From the relation between pressure- and density-perturbations we obtain
\begin{equation}
\d P = \rho \, \d\mut = \cs^2\, \d\rho\,,\implies
n \,\d\mut = \cs^2 \,\d n \,.
\label{eq:compress}
\end{equation}
We write the local sound speed $\cs(r)$ as
\begin{equation}
\cs(r) = \lambda(r)\, \co\,,
\end{equation}
where $\lambda(r)$ is a function of order unity in the bulk of the
star, but which usually vanishes at the stellar surface. As $\d\mut$
has the dimensions of a velocity squared, the relation
(\ref{eq:compress}) takes the following form in natural units: 
\begin{equation}
n\, \d\mut = \lambda^2(r)\, \d n\,,
\end{equation}
so $\d\mut$ and $\d n$ are seen to be of the same order except close
to the surface if \mb{$\lambda\rightarrow 0$}. In this system of units, the
perturbation equations can now be written as 
\begin{eqnarray}
\omh\, \lambda^{-2}\,i (1 + {m \zeta})\,n \,\d\mut + \vnabla\cdot\left( n \, \d\vv \right)&=& 0\,,
\label{eq:ConsAnel}\\
\omh \,\left[i(1+{m \zeta})\d\vv + {2 \zeta}\vz\times\d\vv \right]+ \vnabla \left(
  \d\mut + \d\phi \right) &=& 0\,. 
\label{eq:EulerAnel}
\end{eqnarray}
We restrict ourselves to modes that have low frequencies compared to
the sound-crossing frequency $\co/R$, so we assume
\begin{equation}
\omh \ll 1\,.
\end{equation}
It is straightforward to see from (\ref{eq:EulerAnel}) that
\begin{equation}
\d\mut = \O(\omh)\,,
\end{equation}
and therefore (\ref{eq:ConsAnel}) yields
\begin{equation}
\vnabla\cdot\left( n\, \d\vv \right) = \O(\omh^2\,\lambda^{-2})\,.
\label{eq:ConsAnel1}
\end{equation}
In the bulk of the star, where $\lambda\sim\O(1)$, we
can therefore neglect the density variation $\d n$ in the conservation
equation, leading to an error of order $\O(\omh^2)$. However, in the boundary
layer characterized by \mb{$\lambda\sim\O(\omh)$}, i.e. in the region
where the local sound-speed is of order $\cs\sim \om\,R$, the error of 
neglecting the compressibility of the matter becomes large. 
Nevertheless, the overall quality of the approximation is generally
very good [see \citet{dintrans01:_anelastic}], provided this surface
boundary layer 
is sufficiently thin, but we might expect the surface boundary 
conditions to be modified. This is indeed the case, as
(\ref{eq:ConsAnel1}) now entails the regularity condition \mb{$\left.\d v^r\right|_{r=R} = 0$}
for stellar modes with $\rho\rightarrow 0$ at the surface.  Therefore the surface
displacement is necessarily zero in the anelastic approximation, which
filters out any surface waves.
Another consequence of this approximation is seen by taking the curl
of (\ref{eq:EulerAnel}), which effectively eliminates the potentials
$\d\mut$ and $\d\phi$ from the system of equations. The velocity
perturbation is therefore independent of the pressure- and
gravitational perturbation, which can both be determined a-posteriori
from the solution and the remaining component of the Euler equation. 
The eigenmode solution is therefore independent of all potential
perturbations, $\d\phi$,  $\d P$ (or equivalently $\d\mut$). Although
these perturbations were not assumed to be zero,  they are now 
``slaved'' to the velocity perturbation.

\section{The explicit oscillation equations}
\label{sec:ExplicitEquations}

The general system of equations (\ref{eq:final1})--~(\ref{eq:final4})
for the eigenmode problem together with the definitions in
Sect.~\ref{sec:NumericalResults} can be written in the explicit form 
\begin{eqnarray}
r \, W_\n^{l'} + \left( 1+ r {\rho_\n' \over \rho_\n}\right) W_\n^l - l (l+1)\, V_\n^l  = 0\,,
\label{eq:LSBfirst}\\[0.2cm]
r \, W_\p^{l'} + \left( 1+ r {\rho_\p' \over \rho_\p}\right) W_\p^l -
l (l+1)\, V_\p^l  = 0\,,
\end{eqnarray}
\begin{eqnarray}
(l-1) Q_l \,U_\n^{l-1} - (l+2) Q_{l+1}\, U_\n^{l+1} + m \,V_\n^l \nonumber\\
- (1-\epsn)\gamn m \R\, W_\n^l - \epsn\gamn m \R \, W_\p^l + r\, \psih^{l'}_\n \nonumber\\
= \kap_0\, \left[ (1-\epsn)\gamn \, W_\n^l + \epsn\gamn W_\p^l \right]\,,
\end{eqnarray}
\begin{eqnarray}  
(l-1)Q_l \, U_\p^{l-1} - (l+2)Q_{l+1}\, U_\p^{l+1} + m V_\p^l + r\psih_\p^{l'} \nonumber\\
= \kap_0 \left[ (1-\epsp)\gamp \, W_\p^l + \epsp \gamp \, W_\n^l \right]\,,
\end{eqnarray}
\begin{eqnarray}
(l^2 - 1)Q_l U_\n^{l-1} + l(l+2) Q_{l+1}\, U_\n^{l+1} \nonumber\\
+ \left\{m - l (l+1) (1-\epsn)\gamn m \R \right\}\, V_\n^l \nonumber\\
- l (l+1)\epsn \gamn m \R V_\p^l + m W_\n^l + l (l+1) \psih_\n^l \nonumber\\
= \kap_0 \left[l (l+1) (1-\epsn)\gamn V_\n^l + l (l+1)\epsn \gamn \,
  V_\p^l \right]\,,
\end{eqnarray}
\begin{eqnarray}
(l^2 -1)Q_l \, U_\p^{l-1} + l (l+2) Q_{l+1}\, U_\p^{l+1} \nonumber\\
+ m\, V_\p^l + m\, W_\p^l + l (l+1)\, \psih_\p^l\nonumber\\
= \kap_0 \left[l (l+1)(1-\epsp)\gamp \, V_\p^l + l (l+1)\epsp \gamp \, V_\n^l \right]\,,
\end{eqnarray}
\begin{eqnarray}
\left\{m - l (l+1)(1-\epsn)\gamn m \R \right\}\, U_\n^l \nonumber\\
- l (l+1) \epsn \gamn m \R \, U_\p^l \nonumber\\
+ (l^2-1) Q_l \, V_\n^{l-1} + l (l+2) Q_{l+1} \, V_\n^{l+1} \nonumber\\
- (l+1)Q_l \, W_\n^{l-1} + l Q_{l+1}\, W_\n^{l+1} \nonumber\\
= \kap_0 \left[l (l+1)(1-\epsn)\gamn\, U_\n^l + l (l+1) \epsn \gamn \,U_\p^l \right]\,,
\end{eqnarray}
\begin{eqnarray}
m\, U_\p^l + (l^2 -1)Q_l \, V_\p^{l-1} + l (l+2) Q_{l+1}\, V_\p^{l+1} - \nonumber\\
- (l+1)Q_l \, W_\p^{l-1} + l Q_{l+1}\, W_\p^{l+1} = \nonumber\\
\kap_0 \left[l (l+1) (1-\epsp)\gamp \, U_\p^l + l (l+1) \epsp \gamp \, U_\n^l \right]
\label{eq:LSBlast}
\end{eqnarray}

\end{document}